\documentclass[12pt]{iopart}
\usepackage{iopams} 
\usepackage{times} 
\newcommand{\text}[1]{\mbox{#1}}
\begin{document}
\title[]{1, 2, and 6 qubits, and the Ramanujan-Nagell theorem}
\author{Yaroslav Pavlyukh}
\address{Institut f\"{u}r Physik, Martin-Luther-Universit\"{a}t
  Halle-Wittenberg, Heinrich-Damerow-Str. 4, 06120 Halle, Germany}
\ead{yaroslav.pavlyukh@physik.uni-halle.de}
\author{A.\ R.\  P. Rau} 
\address{Department of Physics and Astronomy, Louisiana State University,
Baton Rouge, LA 70803, USA}
\ead{arau@phys.lsu.edu}
\begin{abstract}
A conjecture of Ramanujan that was later proved by Nagell is used to show on the basis of
matching dimensions that only three $n$-qubit systems, for $n=1, 2, 6$, can share an
isomorphism of their symmetry groups with the rotation group of corresponding dimensions
$3, 6, 91$. Topological analysis, however, rules out the last possibility.
\end{abstract}
\maketitle
Consideration of a single qubit in terms of rotations in three dimensions is
familiar from the earliest days of nuclear magnetic resonance (nmr) and the use of such a
``Bloch sphere" recognized widely as of great value both in nmr and its applications and
in quantum information
\cite{nielsen_quantum_1998,sakurai_modern_1994,ernst_priciples_1987}. Equally, in dealing
with two qubits and their algebra of 15 generators of $SU(4)$, the correspondence to the
15 antisymmetric matrix operators of the six-dimensional rotation group $SO(6)$, with
products of the two sets of Pauli matrices for the two qubits rearranged in such a $6
\times 6$ antisymmetric array has proved useful (see Eq.(B.1) of
\cite{uskov_geometric_2008}). In particular, a ten-dimensional subgroup $SO(5)$ of both
plays an important role in several logic gates and in four-level systems in molecular
physics and quantum optics \cite{uskov_geometric_2008}. Already current
experiments~\cite{blatt_qubit_2011} demonstrate the feasibility of creating multi-qubit states
thus posing a question on whether such parallels can be exploited for even higher
dimensional Lie groups.  The aim of the present contribution is to provide both the
dimensional and topological arguments that these are the only examples of accidental
homeomorphisms between the qubit symmetry groups and the orthogonal groups. In the course
of presentation we will refer to numerous facts from the corresponding
disciplines. Therefore, our exposition will be accompanied by extensive references to
mathematical texts for the interested reader.
\paragraph{Dimensions of unitary and orthogonal groups} 
It is a curious fact that the dimension of the symmetry group of $n$ qubits, $SU(2^n)$, is
of the form of a Mersenne number $2^{2n}-1$. Equally striking even if obvious is the
connection between the mathematical/geometrical concept of triangular numbers and the
number $N(N-1)/2$ of antisymmetric generators that define rotations in $N$-dimensional
space, these $SO(N)$ groups being central to physics.
For the number of generators of $SU(2^n)$ and $SO(N)$ to coincide, we must have
\begin{equation}
N^2-N=2(2^{2n}-1).
\label{eqn1}
\end{equation}
This quadratic equation in $N$ has integer solutions only when the discriminant is a
perfect square, that is, when
\begin{equation}
2^{2n+3}=k^2+7,
\label{eqn2}
\end{equation}
with $k$ an integer.

One hundred years ago in 1913, the mathematical genius Ramanujan, with his gift for number
patterns, conjectured that only a few Mersenne numbers (of the form $2^b -1$ with $b$ a
positive integer) are triangular \cite{ramanujan_question_1913}. This corresponds to the
statement that the Diophantine equation $2^b=k^2+7$ has only a finite number of solutions
among $b$ and $k$ integers. This was proved in 1948 by the mathematician
T. Nagell\footnote{The original proof published in \cite{nagel_1948} is hard to access,
  therefore readers should consult for instance \cite{mollin_advanced_2010}.}, that
such solutions exist only for the five conjectured values, $b=3, 4, 5, 7, 15$ and,
correspondingly, $k=1, 3, 5, 11, 181$ \cite{nagel_1948,mollin_advanced_2010}.

Drawing on the Ramanujan-Nagell theorem, and setting $n=(b-3)/2$, we can conclude that the
only physically relevant values for $n$ qubits are the ones corresponding to the last
three of their numbers, namely, $n=1$ with $(b=5, k=5)$, $n=2$ with $(b=7, k=11)$, and
$n=6$ with $(b=15, k=181)$. Correspondingly, the values of $N$ are $N=(k+1)/2=3, 6,
91$. Thus, three-dimensional rotations for a single qubit, six-dimensional rotations for a
pair of qubits and 91-dimensional rotations for a set of six qubits! But, in the next
section, we show that this last possible homeomorphism can also be ruled out under further
analysis.
\paragraph{Topological analysis}
It is sufficient to consider the simplest topological invariants
(see~\cite{dieudonne_history_2009} on historical perspective on terms mentioned in this
manuscript) the \emph{Betti numbers}, to see that $SU(2^6)$ and $SO(91)$, even though
having the same dimension, differ topologically. We will also operate with the
\emph{Poincar{\'e} polynomial} defined for a $z$-dimensional manifold $M$ as
\[
P(M,t)=\sum_{q=0}^z b_q(M)t^q,
\]
where $b_q$ are the Betti numbers, i.e. the \emph{ranks} (i.e. the number of elements of
  its generating set) of the homology groups $H_q(M,\mathbb{Z})$ of the manifold (very
  transparent introduction of homology groups can be found
  at~\cite{schwartz_topology_2010}). By the theorem of C.~Chevalley (see for example
  \cite{rosenfeld_geometry_1997}, p.~86) the Poincar{\'e} polynomial of a simply connected
  compact simple real Lie group is
\[
P(t)=\prod_i(t^{2\alpha_i+1}+1),
\]
where $\alpha_i$ are the exponents of the simple Lie group. The special unitary groups
$SU(n+1)$ belong to the class $A_n$ (in E.~Cartan
classification~\cite{fuchs_symmetries_1997}) with the exponents $\alpha_i=1,2,\ldots,n$
while the orthogonal group $SO(2n+1)$ belongs to the $B_n$ class with the exponents
$\alpha_i=1,3,\ldots,2n-1$. The Poincar{\'e} polynomials given by:
\begin{eqnarray*}
P(SU(64),t)&=&(1+t^3)(1+t^5)\cdots(1+t^{127}),\\
P(SO(91),t)&=&(1+t^3)(1+t^7)\cdots(1+t^{179})
\end{eqnarray*}
are, thus, different. This demonstrates the topological inequivalence of the two
groups. The theorem of Chevalley applies to Betti numbers defined as ranks of homology
groups with coefficients in $\mathbb{Z}$. Especially for $SO(n)$ they are difficult to
compute. Going to $\mathbb{Z}_2$ presents a simplification and can also be used for
establishing the inequivalence of the two groups. Detailed computation of
$H_*(SO(n),\mathbb{Z})$ and $H_*(SO(n),\mathbb{Z}_2)$ can be found in
\cite{hatcher_algebraic_2002} (see Sec.~3.D).

Since the exponents are of algebraic nature, this consideration of Betti numbers is also
an algebraic argument. From the above argument, one can also easily see that $SU(2)\cong
SO(3)$ is the only example of homeomorphism between the $A_n$ and $B_n$ classes, the
corresponding Poincar{\'e} polynomial being $(1+t^3)$. Similarly, $SU(4)\cong SO(6)$
groups represent the single homeomorphism between the $A_n$ and $D_n$ classes, and the
Poincar{\'e} polynomial is $(1+t^3)(1+t^5)(1+t^7)$.

Summarizing, we have presented a number-theoretic argument which limits the number of possible
homeomorphisms between the $n$-qubit symmetry groups and the special orthogonal
groups. Furthermore, we showed that $SU(2)\cong SO(3)$ and $SU(4)\cong SO(6)$ are the only
homeomorphisms of this kind. Since our reasoning rests on facts from algebra and
topology, we would like to demonstrate it in a more physically transparent way.
\paragraph{Geometric explanation} 
It is desirable to have a simple explanation of the aforementioned fact assuming just
basic knowledge of algebraic topology: we compare the homotopy groups\footnote[1]{The
$k$-th homotopy group $\pi_k(M)$ of the manifold $M$ is the group of continuous mappings
of a $k$-sphere $S^k$ into $M$. For $f,g\in \pi_k(M)$ one defines the group operation
$h=f+g$ as a continuous map from a $S^k$ into the \emph{wedge
sum}~\cite{hatcher_algebraic_2002} of two spheres, i.e. $S^k\rightarrow S^k\vee S^k$
followed by the $f$, $g$ map from each of the two spheres into $M$, respectively.} of the
two spaces. The basis for this comparison is the Bott periodicity
theorem~\cite{bott_periodicity_1970} which applies to \emph{stable} homotopy groups of Lie
groups. For $SO(n)$ and $SU(n)$ the homotopy groups are independent of $n$ if $n$ is
sufficiently large. This can be seen by constructing a \emph{fibration} arising from the
Lie group action:
\[
SU(n-1)\rightarrow SU(n)\rightarrow S^{2n-1},
\]
where the base is the coset space $S^{2n-1}\cong
SU(n)/SU(n-1)$~\cite{schwartz_topology_2010}.  

To see this, we recall that $SU(n)$ acts transitively\footnote{The action of a group $G$
on the space $M$ is called transitive if the orbit of any element $x\in M$ is the whole
space.} on the space $\mathbb C^n$ preserving the norm
$|z_1|^2+\cdots+|z_n|^2=1$. Physically this means that the propagator preserves the
wave-function normalization and each quantum state $\psi_p$ can be labeled by a point on
the sphere $p\in S^{2n-1}$. The stabilizer of this point, i.e. all the transformations
leaving the point unchanged,
\[
\hat U_p\psi_p=\psi_p,
\]
is isomorphic to $SU(n-1)$. This is obvious because $\hat U_p$ performs a transformation
on $\mathbb C^{n-1}\subset \mathbb C^n$, the subspace orthogonal to $p$. We can use now
the exact homotopy sequence to compute homotopy groups of a \emph{principal bundle}
$E(F,M)$, where $F$ is the fiber and $M$ is the base manifold of the fibration (p. 76 of
\cite{josi_lie_1998}):
\[
\ldots\rightarrow\pi_k(F)\rightarrow\pi_k(E)\rightarrow\pi_k(M)\rightarrow\pi_{k-1}(F)\ldots 
\]
Since we know that $\pi_k(S^n)=0$ for $k<n$ we have 
\[
0\rightarrow\pi_k(SU(n))\rightarrow\pi_k(SU(n+1))\rightarrow 0, \quad \text{for}\quad k<2n
\]
and analogically for the 
\[
SO(n-1)\rightarrow SO(n)\rightarrow S^{n-1},
\]
fibration:
\[
0\rightarrow\pi_k(SO(n))\rightarrow\pi_k(SO(n+1))\rightarrow 0, \quad \text{for}\quad k<n-1.
\]
Thus, for $k\le2n-1$ for $SU(n)$ and for $k\le n-2$ for $SO(n)$ we are in stable homotopy
range, i.e. the homotopy groups are independent of $n$: $\pi_k(SU(n))=\pi_k(SU)$ and
$\pi_k(SO(n))=\pi_k(SO)$. This means that up to $k=90$ all homotopies for the two groups
of interest are determined by the Bott periodicity:
\begin{eqnarray*}
\pi_k(U)&=&\pi_{k+2}(U),\\
\pi_k(O)&=&\pi_{k+4}(Sp),\\
\pi_k(Sp)&=&\pi_{k+4}(O),
\end{eqnarray*}
where $U$, $O$ and $Sp$ are the unitary, orthogonal and symplectic groups,
respectively. Already from different periodicities one can conclude that the considered
spaces are topologically distinct. Making use of well known results
\begin{eqnarray*}
0,\,\mathbb{Z}&\quad\text{for}\quad\pi_i(U),&\quad i=0,1,\\
0,\,0,\,0,\mathbb{Z},\,\mathbb{Z}_2,\,\mathbb{Z}_2,\,0,\mathbb{Z}&
\quad\text{for}\quad\pi_i(Sp),&\quad i=0,\ldots,7,\\
\mathbb{Z}_2,\,\mathbb{Z}_2,\,0,\mathbb{Z},\,0,\,0,\,0,\mathbb{Z}&
\quad\text{for}\quad\pi_i(O),&\quad i=0,\ldots,7.
\end{eqnarray*}
and the fact that $\pi_k(U)=\pi_k(SU)$ for $k\ge2$ and $\pi_k(O)=\pi_k(SO)$ we observe that
$\pi_5$ groups are distinct:
\begin{eqnarray*}
\pi_5(SU(64))=\pi_5(SU)=\pi_1(SU)&=\mathbb{Z},\\
\pi_5(SO(91))=\pi_5(SO)=\pi_1(Sp)&=0.
\end{eqnarray*}
It should be noted that there is no contradiction for $SO(3)$ and $SO(6)$ groups: they
fall outside the stable homotopy range for the $\pi_5$ homotopy group.  Finally we remark that
the Bott periodicity, although formulated in pure topological terms, follows from the
underlying algebraic structure of the Lie groups.  
\paragraph{Conclusions} We have demonstrated on the basis of the Ramanujan-Nagell theorem
that $SU(2)\cong SO(3)$ and $SU(4)\cong SO(6)$ are the only accidental homeomorphisms
between the symmetry groups of qubits and the orthogonal groups. Interestingly, another
pair, $SU(2^6)$ and $SO(91)$ groups, also coincide in dimension but they are topologically
and algebraically distinct. Even though the homeomorphism is not present, the 6-qubit
system can efficiently be treated for the purpose of quantum computation by the Cartan
decomposition~\cite{khaneja_cartan_2001}.

Finally, we note a possible further relevance of our observation stemming from the fact
that subgroups of the full $SU(2^n)$ are often important in physical applications of $n$
qubits as in our example \cite{uskov_geometric_2008} of $SO(5)$ for $n=2$. Relaxing the
Ramanujan focus on Mersenne numbers, were we to consider a more general $SU(d)$ with $d$
an integer smaller than $2^n$, the condition in Eq.~(\ref{eqn2}) becomes $8d^2=k^2+7$
which has many more solutions:
\begin{eqnarray*}
n_{i+1}=3n_i+k_i,\\
k_{i+1}=8n_i+3k_i,
\end{eqnarray*}
with initial values $(n_0,k_0)=(1,\pm1)$.  One example is $d=11, k=31, N=16$. Furthermore,
 in a $n$-qubit system, even though the full symmetry group $SU(2^n)$ may not be
 isomorphic to a rotation group, there is a number of rotational subgroups:
\[
SU(2^n)\supset SO(2^n)\supset SO(2^{n-1})\ldots 
\] 
Such correspondences may be exploited for other systems of multiple qubits.
\section*{Acknowledgments}
One of us (ARPR) thanks the Alexander von Humboldt Stiftung for support and Drs. Gernot
Alber and Jamal Berakdar for their hospitality during the course of this work.

\end{document}